\documentclass[twocolumn,letter]{jpsj3_pdftex}
\usepackage{amsmath}
\usepackage{amssymb}
\usepackage{bm}
\usepackage{color}
\usepackage{graphicx}
\allowdisplaybreaks

\title{%
Magnetization Step in Spatially Distorted Heisenberg Kagom\'{e}
Antiferromagnets
}%
\author{%
Ryui \textsc{Kaneko},
Takahiro \textsc{Misawa},
and
Masatoshi \textsc{Imada}
}%
\inst{%
Department of Applied Physics, University of Tokyo, 7-3-1 Hongo,
Bunkyo-ku, Tokyo 113-8656
}%

\recdate{\today}

\abst{%
Motivated by a recent experiment on volborthite, a typical spin-$1/2$
antiferromagnet with a kagom\'{e} lattice structure, we study the magnetization
process of a classical Heisenberg model on a spatially distorted kagom\'{e}
lattice using the Monte Carlo (MC) method.  We find a distortion-induced
magnetization step at low temperatures and low magnetic fields.
The magnitude of this step is given by $\Delta
m_z=\left|1-\alpha\right|/3\alpha$ at zero temperature, where $\alpha$
denotes the spatial anisotropy in exchange constants.  The magnetization
step signals a first-order transition at low temperatures, between two
phases distinguished by distinct and well-developed short-range spin
correlations, one characterized by spin alignment of a local $120^{\circ}$
structure with a $\sqrt{3}\times\sqrt{3}$ period, and the other by a
partially spin-flopped structure.  We point out the relevance of our
results to the unconventional steps observed in volborthite.
}%
\kword{%
volborthite, magnetization step, geometrical frustration, kagom\'{e}
lattice, Heisenberg model
}%

\begin{document}

\maketitle

In antiferromagnets on the kagom\'{e} lattice, geometrical frustration
effects are believed to suppress the conventional magnetic long-range
order and induce a large degeneracy of the ground state.  It opens a
possibility of realizing spin-liquid states, which are spin analogues of
liquids~\cite{Eur.Phys.J.B.2.501}.  Volborthite
Cu$_3$V$_2$O$_7$(OH)$_2\cdot$2H$_2$O is known as a nearly ideal
spin-$1/2$ kagom\'{e} antiferromagnet~\cite{JSSC.85.220}.  Measurements
of the magnetization and specific heat for this compound indicate the
absence of the conventional magnetic long-range order down to
50 mK~\cite{JPSJ.78.043704, JPSJ.70.3377}.  Therefore, it is proposed
that this material offers evidence of the quantum spin-liquid state
in nature~\cite{JPSJ.70.3377}.

Recent studies on volborthite have revealed the existence of
three unconventional steps in the magnetization
curve~\cite{JPSJ.78.043704}.  These steps are not anticipated in the
magnetization process of the isotropic kagom\'{e} Heisenberg model in the
literature~\cite{PhysRevLett.88.057204,JPSJ.70.3673} and their origin
is not yet understood.  A MC study of the classical Heisenberg
model has clarified the existence of the magnetization plateau at
one-third of the saturation magnetization ($m_{\mathrm{sat}}/3$) at
finite temperatures, interpreted as the order-by-disorder
effect~\cite{PhysRevLett.88.057204}.  Even for the quantum Heisenberg
model ($S=1/2$ and $1$), an exact diagonalization study has shown a similar
magnetization plateau at zero temperature~\cite{JPSJ.70.3673}.  However,
the magnetization step has not been found in the isotropic
kagom\'{e} Heisenberg model.

In this Letter, to shed light on the origin of the unconventional steps
observed in volborthite, we calculate the magnetization curve in the
antiferromagnetic classical Heisenberg model on the spatially distorted
kagom\'{e} lattice by using the MC method.  In our viewpoint, structural
distortion, which inevitably exists in volborthite~\cite{JPSJ.70.3377},
holds the key to understanding the magnetization step.  This structural distortion
induces a spatial anisotropy in exchange interactions.  Therefore, as a
simple model, we employ an anisotropic ($J_1$-$J_2$) kagom\'{e}
Heisenberg model to describe the magnetization process observed in
volborthite.  We find distortion-induced first-order phase transitions
at low temperatures and low magnetic fields.  In the vicinity of this
first-order phase transition, step-like behaviors of the magnetization
curve are found and spin structure factors change drastically from local
$120^{\circ}$ structures to spin-flopped structures.  We also determine
the $h$-$T$ phase diagram for the spatially anisotropic kagom\'{e}
Heisenberg model.

\begin{figure}[hbpt]
\centering%
\includegraphics[width=.43\textwidth]{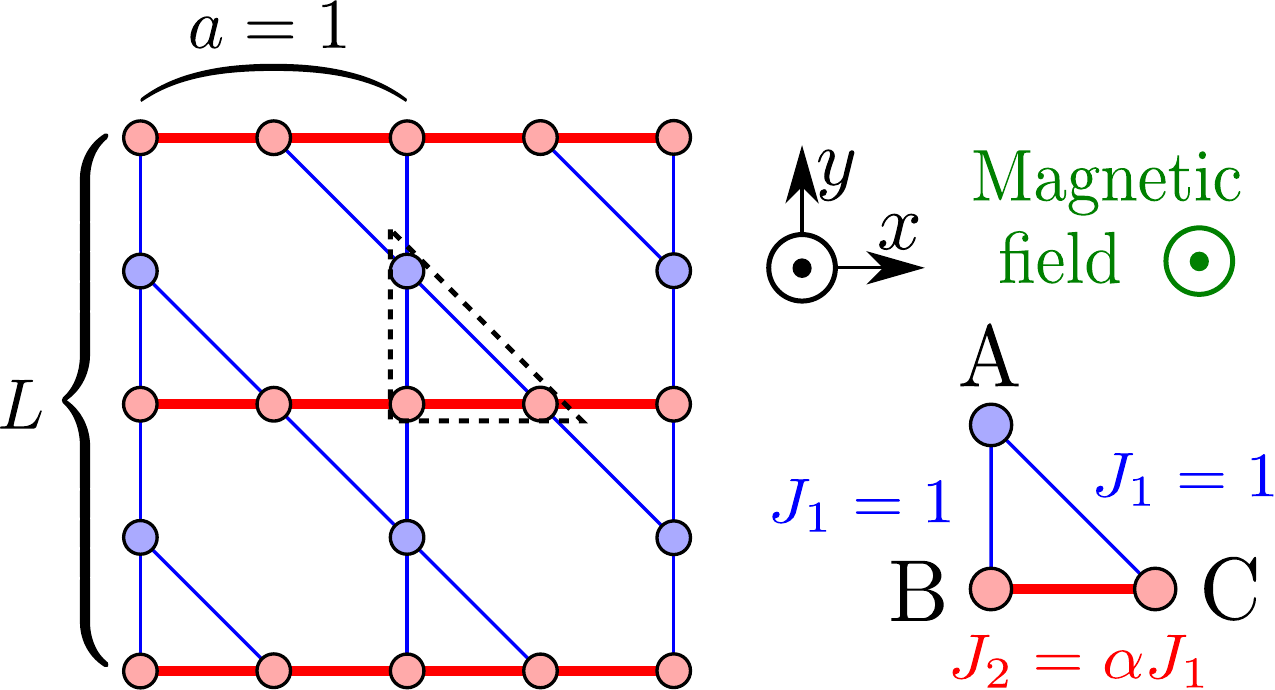}%
\caption{%
(Color online)
An $L\times L$ ($L=2a$) distorted kagom\'{e} lattice.  Lattice sites
labeled A, B, and C constitute a unit cell (dashed-line triangle).
The number of spins is $N=3L^2$.  Bonds along two directions have the
exchange constant $J_1$ (thin blue lines), while the bond along the
third direction has the exchange constant $J_2=\alpha J_1$ (thick red
lines).  For volborthite, it is plausible that $\alpha>1$.
}
\label{fig:1}
\end{figure}

Our model Hamiltonian is defined in the form
\begin{equation}
 H =
   J_{1} \sum_{\left<i,j\right>_{1}}
   \bm{S}_{i} \cdot \bm{S}_{j}
 + J_{2} \sum_{\left<i,j\right>_{2}}
   \bm{S}_{i} \cdot \bm{S}_{j}
 - h \sum_{i} S_{i}^{z},
\end{equation}
where $\bm{S}_{i}$ denotes the Heisenberg classical spin, $h$ denotes the
external magnetic field, and $\left<i,j\right>_{1}$ and
$\left<i,j\right>_{2}$ denote nearest-neighbor sites on AB/AC chains and
BC chains, respectively.  In the kagom\'{e} lattice, the unit cell contains
three sites A, B, and C, as shown in Fig.~\ref{fig:1}.  The spatial
anisotropy induces different antiferromagnetic couplings on each
triangle: two $J_1$ bonds and one $J_2$ bond.  For simplicity, we set
$J_{1}=1$ and $J_{2}=\alpha J_{1}$.  We simplify the Hamiltonian as
\begin{eqnarray}
  H = \sum_{\mathrm{triangles}}
 \bigg[
  \left(
     \bm{S}_{\mathrm{A}} \cdot \bm{S}_{\mathrm{B}}
   + \alpha \bm{S}_{\mathrm{B}} \cdot \bm{S}_{\mathrm{C}}
   + \bm{S}_{\mathrm{C}} \cdot \bm{S}_{\mathrm{A}}
  \right)\nonumber\\
   -\frac{h}{2}
    \left(
       S_{\mathrm{A}}^{z}
     + S_{\mathrm{B}}^{z}
     + S_{\mathrm{C}}^{z}
    \right)
 \bigg].
\end{eqnarray}
The summations are taken over upward and downward triangles.  The
lattice has no distortion when $\alpha=1$.  There are two simple limits
for this model.  For $\alpha\rightarrow0$, the lattice becomes a
decorated square lattice, with additional sites at the midpoints of
square lattice edges.  In this limit, the ground state is ferrimagnetic.
For $\alpha\rightarrow\infty$, the lattice becomes isolated
antiferromagnetic chains and free spins, and is equivalent to the
one-dimensional lattice.  In volborthite, we infer that $\alpha$ is
larger than unity from a comparison of bond lengths.  The magnitude of the
anisotropy is estimated to be less than $20\%$ from theoretical
analysis~\cite{arXiv:0707.4264}.  We consider both kinds of anisotropy,
especially $\alpha=1.05$ and $\alpha=0.95$, but perform detailed
analyses of the $\alpha=1.05$ case.

We clarify thermodynamic properties of this model by using MC
simulations.  The Metropolis algorithm is employed.  We start the
simulations from random initial states.  Spins are updated in the
sequential order on the lattice.  We also perform one over-relaxation
update per MC step to accelerate efficient
sampling~\cite{PhysRevD.36.515}.  At every temperature and field,
$5\times 10^4$ MC steps are taken for equilibration.  These are
followed by measurements conducted for more than $5\times 10^5$ MC steps.
Finally, the results are averaged over more than 8 independent
runs to estimate statistical errors.  At low temperatures $T\le 0.01$,
we use the exchange MC algorithm~\cite{JPSJ.65.1604}.  Simulations are conducted
for system sizes up to $L=72$ ($N=15552$) under the periodic boundary
conditions.  We set the Boltzmann constant $k_{\mathrm{B}}=1$.

\begin{figure}[hbpt]
\centering%
\includegraphics[width=.43\textwidth]{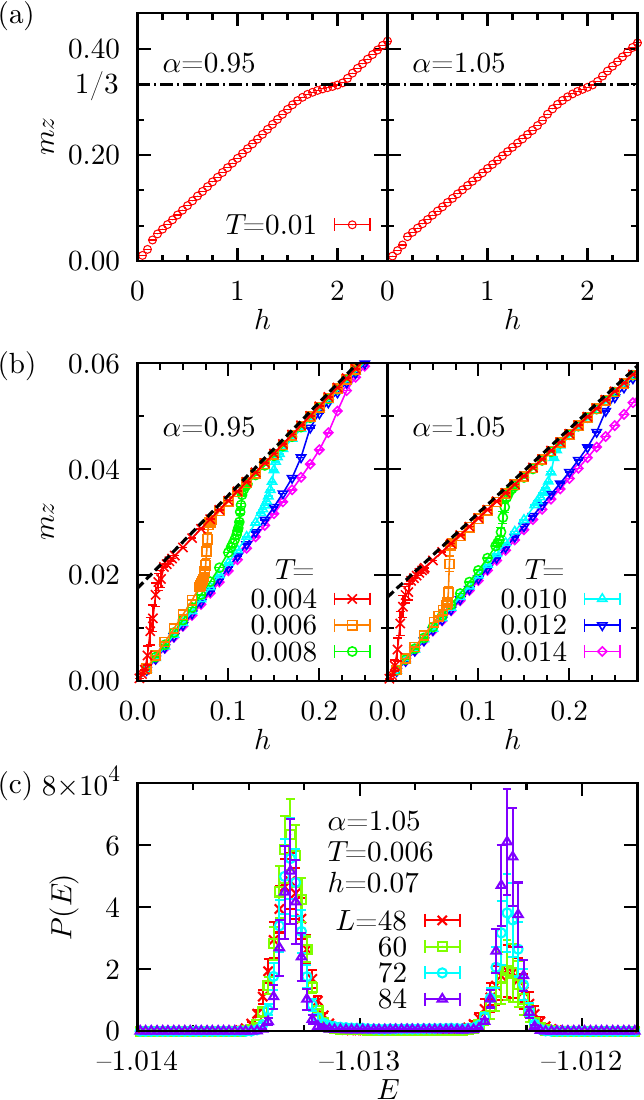}%
\caption{%
(Color online)
(a) Magnetization curves at $T=0.01$ for $\alpha=1.05$ and
$\alpha=0.95$.  The system size is $L=36$.  Statistical errors are
smaller than the symbol size.  The dashed-dotted line denotes one third
of the saturation magnetization.  For both anisotropies, magnetization
plateaus exist at $h\simeq 2$.
(b) Magnetization curves at low magnetic fields.  Step-like behaviors in
the magnetization curve are observed for both $\alpha=1.05$ and
$\alpha=0.95$.  The system size is $L=36$.  The step shows first-order
transitions below the critical temperature $T_c\simeq 0.007$.  At
$T>T_c$, the step vanishes and turns into a crossover.  We have observed
a magnetization hysteresis for $T\le 0.008$.  Statistical errors are
smaller than the symbol size at $T=0.008$.  Magnetization curves in the
decreasing field are shown at $T\le 0.006$.  Owing to finite size effects,
a first-order step at $T=0.004$ appears to be rounded.  The dashed line
denotes the magnetization at $T=0$ given by
$m_z=(h+2\left|1-\alpha\right|)/(6\alpha)$. See the text for details.
(c) Energy distribution $P(E)$ at $T=0.006$ and $h=0.07$ for $L=48$, $60$,
$72$, and $84$. The two-peak structure still survives for large systems.
}
\label{fig:2}
\end{figure}

In Fig.~\ref{fig:2}(a), we show the magnetization $m_z$ parallel to applied
fields at several choices of temperatures as a function of magnetic
fields.  For both $\alpha>1$ and $\alpha<1$, we have observed a
magnetization plateau $m_{\mathrm{sat}}/3$ at low temperatures, which
has already been reported by Zhitomirsky for the isotropic ($\alpha=1$)
kagom\'{e} lattice~\cite{PhysRevLett.88.057204}.
Owing to the distortion, the magnetization plateau
becomes much wider.  This tendency is consistent with the results of the
exact diagonalization for a spin-$1/2$ distorted kagom\'{e} Heisenberg
model~\cite{JPSJ.70.3673}.

As shown in Fig.~\ref{fig:2}(b), we have observed step-like behaviors
at low temperatures and low magnetic fields.  To determine whether these
step-like behaviors are the first-order phase transitions in the
thermodynamic limit, we calculate the energy distribution $P(E)$ up to
$L=84$.  If the transition is of the first order, the energy
distribution $P(E)$ should be bimodal at the transition temperature.  As
shown in Fig.~\ref{fig:2}(c), we find a clear two-peak structure around
the transition point.  This result shows evidence for the first-order
phase transitions at low temperatures.  These first-order phase
transitions appear only when the lattice has anisotropies
($\alpha\not=1$).  The transition field increases monotonically with
increasing temperature.  The first-order jump becomes reduced with
increasing temperature and vanishes at a finite-temperature critical
point, as we detail below.

The origin of the first-order phase transitions is well understood by
considering the zero temperature limit.  To analyze the ground-state
properties, we rewrite the Hamiltonian as
\begin{equation}
  H = \sum_{\mathrm{triangles}}
 \left[
  \frac{\alpha}{2}
  \left(
    \bm{S}_{\triangle}
   -\frac{h}{2\alpha}\bm{e}_z
  \right)^2
  + \frac{1-\alpha}{2\alpha}hS_{\mathrm{A}}^z
%
  + \mathrm{const.}
 \right],
\end{equation}
where $\bm{S}_{\triangle} =
\bm{S}_{\mathrm{A}}/\alpha+\bm{S}_{\mathrm{B}}+\bm{S}_{\mathrm{C}}$.
For $h=0$, the energy is minimized with $\bm{S}_{\triangle}=0$.  The
ground state, which satisfies the constraint $\bm{S}_{\triangle}=0$, has
an infinite degeneracy and the magnetization is zero.  For
$0<h<\mathrm{min}(2,4\alpha-2)$, the energy is minimized with
$\bm{S}_{\triangle}=(h/2\alpha)\bm{e}_z$ and
$S_{\mathrm{A}}^z=-\mathrm{sign}(1-\alpha)$, and hence
$S_{\mathrm{B}}^z=S_{\mathrm{C}}^z=
[h+2\,\mathrm{sign}(1-\alpha)]/(4\alpha)$.  Therefore, at zero
temperature, we obtain the magnetization as
\begin{equation}
 m_z
 = \frac{S_\mathrm{A}^z+S_\mathrm{B}^z+S_\mathrm{C}^z}{3}
 = \frac{h+2\left|1-\alpha\right|}{6\alpha}.
\label{Eq:mz}
\end{equation}
Equation (\ref{Eq:mz}) indicates that the magnetization jumps by an
infinitesimal magnetic field at $T=0$.  Although thermal fluctuations
reduce the magnitude of this jump and shift the transition point to
higher fields, this first-order phase transition still survives at
sufficiently low temperatures, as we have already shown by MC
calculations.

From the above analyses, it turns out that the origin of the first-order
phase transition is a level crossing of the ground state.
When the spatial anisotropy exists, the macroscopic degeneracy of
the kagom\'{e} antiferromagnet is partially lifted by an infinitesimal
magnetic field, i.e., under magnetic fields, the ordered vector
becomes $\bm{q}=(0, q_{y})$, where the range of $q_y$ is given by
$-\pi\leq q_{y}\leq\pi$.

To identify the nature of the first-order transition, we calculate at
$T>0$ the spin structure factors defined as
\begin{equation}
 S^{xy}(\bm{q}) = \frac{1}{N^2}\sum_{l,i,j}
 \left<S^x_{li}\cdot S^x_{lj} + S^y_{li}\cdot S^y_{lj}\right>
 e^{i\bm{q}\cdot(\bm{R}_i-\bm{R}_j)},
\end{equation}
where the index $l$ denotes the positions of the three sites on the unit
cell, and $i$, $j$, and $\bm{R}_{i,j}$ denote the coordinates of the
$(i,j)$-th unit cell on the triangular Bravais lattice.
We take the lattice period (a side length of the square in
Fig.~\ref{fig:1}) as the length unit
and hence the interspin distance equals
$1/2$.  The first Brillouin zone is a square bounded by $-\pi\le q_x,
q_y\le \pi$.

\begin{figure}[hbpt]
\centering%
\includegraphics[width=.40\textwidth]{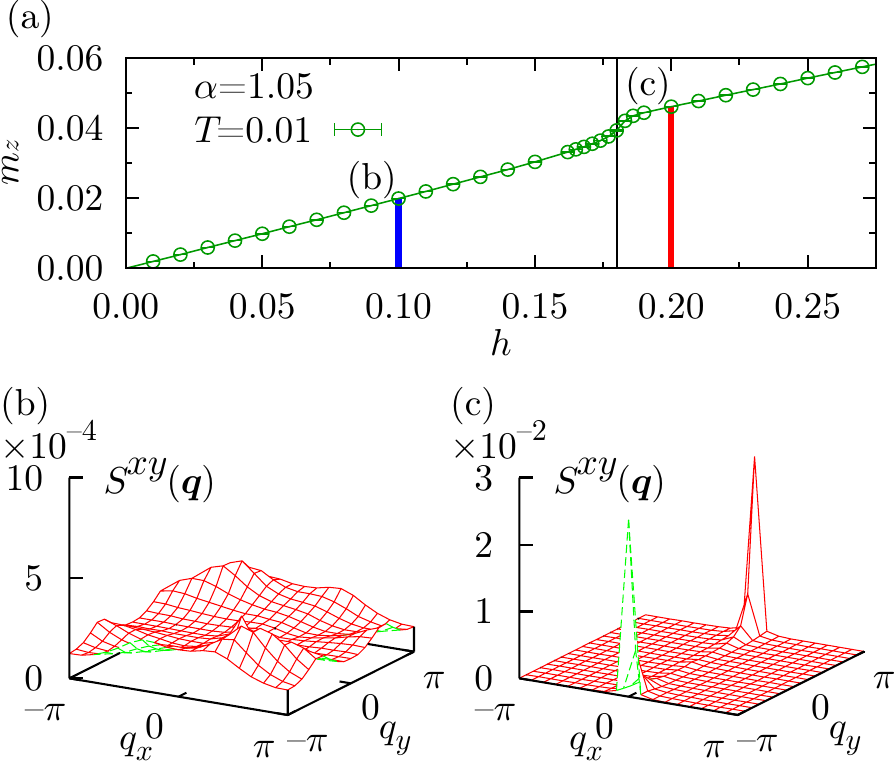}\\[\baselineskip]%
\includegraphics[width=.38\textwidth]{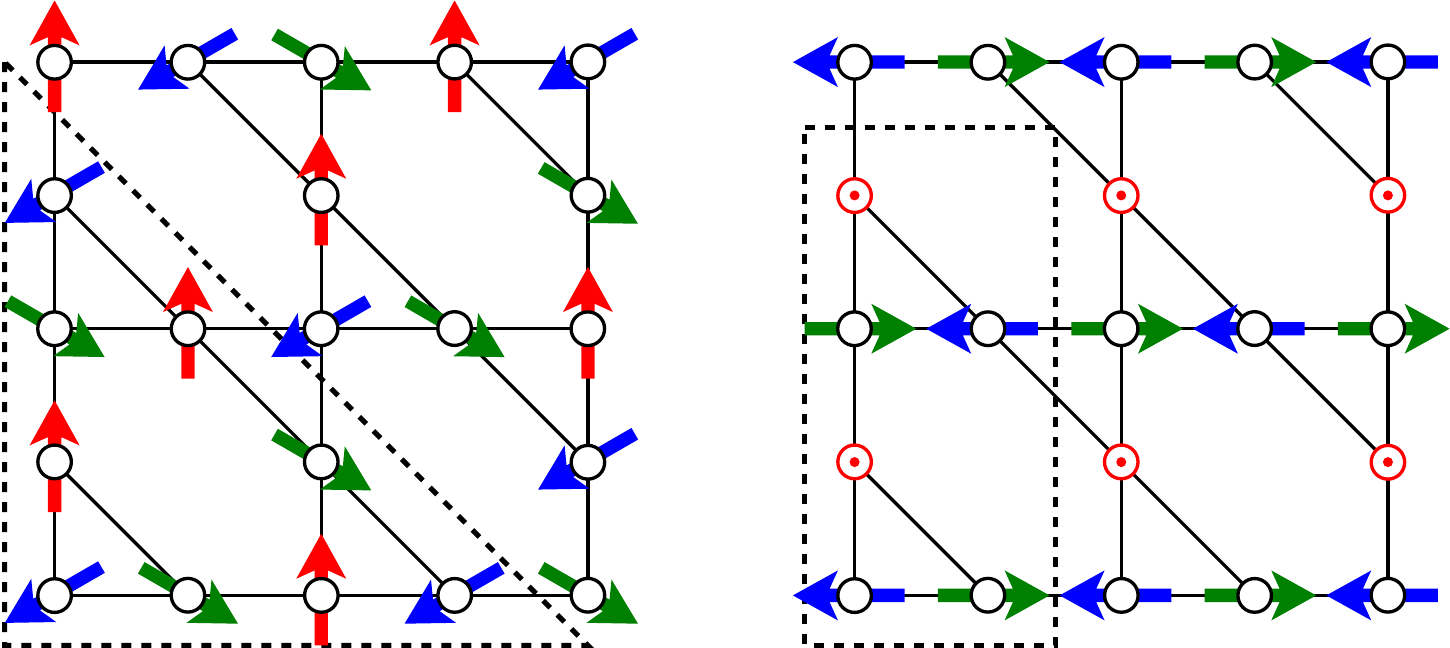}%
\caption{
(Color online)
(a) Magnetization curve around the first-order transition at $T=0.01$.
Statistical errors are smaller than the symbol size.
(b) and (c) Spin structure factors for $\alpha=1.05$, $L=36$ (upper
panel) and corresponding schematic spin configurations in real space
(lower panel) around the first-order transition.
In (b), $h$ is lower than the transition field $h_c\simeq 0.16$
($h=0.1$, $T=0.01$), where the structure factor shows a
$\sqrt{3}\times\sqrt{3}$ pattern.
In (c), $h$ is higher than $h_c$ ($h=0.2$, $T=0.01$), where the
structure factor shows a $\bm{q}=(0, \pm\pi)$ pattern.
The peak values are $S(\bm{q})=3.21(1)\times 10^{-4}$ in (b) and
$S(\bm{q})=2.6(2)\times 10^{-2}$ in (c) (the last digit in the
parentheses indicates the error bar).  In spin structure factors, we
illustrate the regions $-\pi\le q_x\le \pi$ and $-\pi\le q_y\le\pi$ in
the momentum space.  In schematic spin configurations illustrated in the
lower panels of (b) and (c), the unit cells of each pattern are shown by
a dotted line.
}
\label{fig:3}
\end{figure}

As shown in Fig.~\ref{fig:3}, the structure factors $S^{xy}(\bm{q})$
drastically change when we cross the first-order transition point.  At
lower fields~[$h=0.1$, $T=0.01$, see Fig.~\ref{fig:3}~(b)],
$S^{xy}(\bm{q})$ has broad peaks at $\bm{q}=(\pm 2\pi/3,\mp 2\pi/3)$,
which corresponds to a $\sqrt{3}\times\sqrt{3}$ pattern with the
$120^{\circ}$ structure in real space.  At higher fields~[$h=0.2$,
$T=0.01$, see Fig.~\ref{fig:3}(c)], the structure factor has sharp
peaks at $\bm{q}=(0,\pm \pi)$, indicating the real space pattern shown
in the lower panel of Fig.~\ref{fig:3}(c).  In the magnetic structures
in Fig.~\ref{fig:3}(c), spins on A sites align parallel to the magnetic
field indicating a partial spin flip under magnetic fields.  Along a BC
chain, the $xy$ components of spins on the B sites are all parallel to each
other and are antiparallel to all the spins on the C sites.  Along an AB
chain, the spins on the B sites align alternately.  Although the
ground states are degenerate among any ordered vector
$\bm{q}=(0,q_{y})$, thermal fluctuations appear to favor the
$\bm{q}=(0,\pm \pi)$ configuration via the order by a disorder
effect~\cite{PhysRevLett.69.832, JApplPhys.73.5639}.

\begin{figure}[hbpt]
\centering
\includegraphics[width=.43\textwidth]{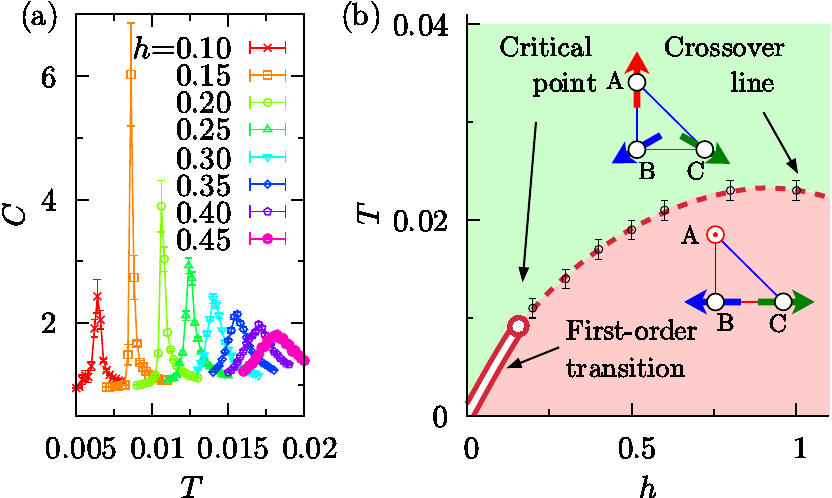}%
\caption{%
(Color online)
(a) Temperature dependence of specific heat at several magnetic fields
which are labeled from $h=0.10$ to $0.45$ starting from the left
($\alpha=1.05$, $L=60$).
(b) Phase diagram of distorted kagom\'{e} lattice ($\alpha=1.05$).  The
first-order phase transition line terminates at the critical point (open
circle).  The dashed line represents a crossover line, which is
characterized by the peak of the specific heat.
}%
\label{fig:4}
\end{figure}

We obtain phase boundaries for $\alpha=1.05$ at lower temperatures and
lower fields $h<2$, as shown in Fig.~\ref{fig:4}.
From the location of the steps, we determine the
first-order transition line.  Because this first-order transition is
characterized by the jump of the magnetization and does not involve any
explicit symmetry breaking, its nature is inferred to be essentially the
same as that of the gas-liquid type.  Namely, the uniform magnetic
susceptibility diverges at the critical end point of this first-order
phase transition as the density fluctuation diverges at the critical end
point of the gas-liquid transition.  Thus, the first-order phase
transition line terminates at the critical point, and a crossover line
appears beyond the critical point.  This crossover line is determined by
the peak of the specific heat and the susceptibility.  Since there is an
exact correspondence between the gas-liquid transition and the
two-dimensional Ising transition,~\cite{book:Goldenfeld} the
universality class of the critical point of this first-order transition
belongs to that of the Ising model.  Therefore, as one approaches the
critical point along the crossover line, the susceptibility diverges as
$\chi_{\mathrm{peak}} \propto |T-T_c|^{-\gamma}$ or $|h-h_c|^{-\gamma}$,
where $\gamma=7/4$ for the Ising universality
class~\cite{book:Goldenfeld}.  By using this relation, we roughly
estimate the location of the critical point as
$(h_c,T_c)=[0.11(1),0.007(1)]$.

Although continuous symmetry breaking is prohibited in two dimensions at
$T\not=0$, a quasi-long-range order may occur in the case of the $XY$
anisotropy~\cite{PhysRevLett.17.1133}.  Because the effective symmetry
of the Heisenberg model becomes of the $XY$ type under the magnetic fields, the
quasi-long-range order of the spin itself~\cite{ZhEkspTeorFiz.59.907,
ZhEkspTeorFiz.61.1144, JPhysC.6.1181, JPhysC.7.1046} or the triatic
order~\cite{PhysRevB.45.7536, PhysRevB.55.11745, PhysRevB.64.134522},
i.e., Berezinskii-Kosterlitz-Thouless (BKT) transition, may occur.
To examine the possibility of the quasi-long-range order, we have
calculated the spin correlation function and estimated the correlation
lengths below the crossover line.  It seems that the spin correlations
along the AB and BC chains decay exponentially for a system size $L\ge 48$
even at the lowest accessible temperature.  Therefore, we tentatively conclude that
no BKT phase of spins exists under magnetic fields, although detailed
studies are left for future studies.  Although the
possibility of the other BKT-type phase transitions is not excluded by
our calculation, such transitions are unlikely because the system
becomes an effectively disconnected one-dimensional BC chain at zero
temperature (see Fig.~\ref{fig:3}) and has no two-dimensional long-range
order. This absence of two dimensional long-range order indicates the
fact that no BKT transition occurs at finite temperatures.

We now examine the relevance of our results to the unconventional steps
observed in volborthite.  In our model, we find one distortion-induced
magnetization step.  This first-order transition is similar to the
step-like behaviors in volborthite, although three steps are observed in
the experiment.  To identify the relationship between the experiment and
the result of our model, we estimate the upper bound of the
magnetization at the transition point.  At $T>0$, the transition occurs
between two nonzero $m_z$'s, namely $m^{+}$ and $m^{-}$, at a transition
field $h_{\mathrm{1st}}>0$.  As can be seen from Fig.~\ref{fig:2}(b), for all
temperatures, $m^{+}$ is nearly the same as that of the zero-temperature
magnetization, i.e.,
$m_z(h=h_{\mathrm{1st}})=[h_{\mathrm{1st}}+2(\alpha-1)]/(6\alpha)$.
Since the distortion is 20\% at most ($\alpha<1.2$) and the transition
occurs at $h_{\mathrm{1st}}<h_c\sim 0.11$, we obtain $m^{+}\lesssim
1/14$.  By considering the experimental magnetization where the three
steps occur ($m_{\mathrm{sat}}/3$, $m_{\mathrm{sat}}/6$, and
$m_{\mathrm{sat}}/45$), if we assign $m_{\mathrm{sat}}/14$ to an upper
bound of one of them, it is likely that the $m_{\mathrm{sat}}/45$ step
corresponds to the present distortion-induced step.  The sudden change
in the spin structure factors from $\bm{q}=(\pm 2\pi/3,\mp 2\pi/3)$ to
$\bm{q}=(0,\pm \pi)$ structures at the first-order transition may be
detected in neutron scattering and nuclear magnetic resonance
experiments.

In the experiment, the transition point does not sensitively depend on
$T$, while our results imply that it should decrease to $h_{\mathrm{1st}}=0$ when
$T\rightarrow 0$. This discrepancy may be due to quantum effects under
which the $\bm{q}=(\pm 2\pi/3,\mp 2\pi/3)$ structure may be stabilized even
at $h\not=0$ and $T=0$.  It is intriguing to identify the critical point
for volborthite and to determine whether the first-order transition survives
under quantum fluctuations.

In summary, by using the MC method, we have clarified the thermodynamic
properties of the classical Heisenberg antiferromagnets on a distorted
kagom\'{e} lattice under magnetic fields. The distortion-induced
magnetization step appears at low temperatures and low magnetic fields.
Estimating the upper bound of the step size, we conclude that this is
the first-order transition induced by spatial anisotropy and
corresponds to the step at the lowest field observed for volborthite.
The spin structure factor shows a sharp change at the transition, which
may be detected experimentally.

Clarifying the origins of the two other steps in the experiment remains for
future studies.  Other interactions in volborthite such as next-nearest
neighbor interactions and Dzyaloshinskii-Moriya interactions may give
rise to additional unconventional phase transitions.  Moreover, our
results are obtained for the classical model while volborthite consists
of quantum spin-$1/2$ spins.  It is an interesting future work to determine whether
quantum fluctuations suppress the first-order transition observed in
the classical system, leading to a crossover with a similarity to the
experiment in which the temperature dependence of the transition field is weak.

\acknowledgements{%
The authors thank Hiroshi Shinaoka,
Youhei Yamaji, and Yukitoshi Motome for fruitful
discussions.  The authors also thank Zenji Hiroi for illuminative
discussions on experimental results.  This work was supported by a
Grant-in-Aid for Scientific Research on Priority Areas (No. 17071003)
from MEXT, Japan.
}



\end{document}